\documentclass[
    ,final            % use final for the camera ready runs
  ]
  {aipproc}

\layoutstyle{6x9}

\def\be{\begin{equation}}
\def\ee{\end{equation}}
\def\beq{\begin{eqnarray}}
\def\eeq{\end{eqnarray}}

\def\IL{\relax{\rm I\kern-.18em L}}

\def\f{\frac}

\begin{document}

\title{Considerations on the excitation of black hole quasinormal modes}

\classification{04.70.-s, 04.30.Db, 04.80.Cc}
\keywords      {black holes, gravitational waves, quasinormal modes}

\author{Emanuele Berti}
{address={McDonnell Center for the Space Sciences, Department of
Physics, Washington University, St.  Louis, Missouri 63130, USA}}

\author{Vitor Cardoso}
{address={Department of Physics and Astronomy, The University of
Mississippi, University, MS 38677-1848, USA}}

\author{Clifford M. Will}
{address={McDonnell Center for the Space Sciences, Department of
Physics, Washington University, St.  Louis, Missouri 63130, USA}}

\begin{abstract}
We provide some considerations on the excitation of black hole
quasinormal modes (QNMs) in different physical scenarios.  Considering
a simple model in which a stream of particles accretes onto a black
hole, we show that resonant QNM excitation by hyperaccretion requires
a significant amount of fine-tuning, and is quite unlikely to occur in
nature. Then we summarize and discuss present estimates of black hole
QNM excitation from gravitational collapse, distorted black holes and
head-on black hole collisions. We emphasize the areas that, in our
opinion, are in urgent need of further investigation from the point of
view of gravitational wave source modeling.
\end{abstract}

\maketitle

A leading candidate source of detectable waves for Earth-based
interferometers such as LIGO, Virgo, GEO600 and TAMA, as well as for
the space-based interferometer LISA, is the inspiral and merger of
binary black holes.  The waveform should comprise three parts, usually
referred to as inspiral, merger and ringdown.  The inspiral waveform,
originating from that part of the decaying orbit leading up to the
innermost stable orbit, can be analyzed using post-Newtonian theory.
Extensive studies of the detectability of this phase of the signal
have been carried out, both for Earth-based \cite{BCV} and for
space-based interferometers \cite{BBW}. The nature of the merger
waveform is largely unknown at present, and is the subject of work in
numerical relativity.

The ringdown waveform originates from the distorted final black hole,
and consists of a superposition of quasi-normal modes (QNMs).  Each
mode has an oscillation frequency and a damping time that are uniquely
determined by the mass $M$ and specific angular momentum $j\equiv
J/M^2$ of the black hole \cite{kokkotas}.  The amplitudes and phases
of the various modes are determined by the specific process that
formed the final hole.

The uniqueness of the modes' frequencies and damping times is directly
related to the ``no hair'' theorem of general relativistic black
holes. Therefore a reliable detection and accurate identification of
QNMs could provide the ``smoking gun'' for black holes and an
important test of general relativity in the strong-field
regime \cite{dreyer}.

The detectability of ringdown waves, and the accuracy with which we
can measure their frequencies and damping times to test the no-hair
theorem, depend mainly on the energy carried by each mode
\cite{BCW}. In turn, the energy distribution depends on the details of
the merger process. Given our poor understanding of the merger phase,
we have at best sketchy information concerning the energy distribution
between different modes.

In the first part of this paper we show by a simple toy model that QNM
excitation by infalling matter requires a significant amount of
fine-tuning, and is quite unlikely in astrophysically realistic
scenarios. In the second part we briefly review present estimates of
QNM excitation in various physical situations, including gravitational
collapse, simulations of single distorted black holes and head-on
collisions of two black holes. This second part is an extended version
of Sec.~VB in \cite{BCW}. Most of our considerations can be applied to
the solar-mass black holes detectable by Earth-based interferometers,
but the main motivation for this short review comes from our study of
the massive black holes detectable by LISA \cite{BCW}. Throughout this
paper we use geometrical units ($G=c=1$).

%%%%%%%%%%%%%%%%%%%%%%%%%%%%%%%%%%%%%%%%%%%%%%%%%%%%%%%%%%%%%%%%%%%%%%%%%%%%%%%
\section{Excitation by infalling matter}
%%%%%%%%%%%%%%%%%%%%%%%%%%%%%%%%%%%%%%%%%%%%%%%%%%%%%%%%%%%%%%%%%%%%%%%%%%%%%%%

Various authors \cite{fhh,AG} suggested that quasinormal ringing could
be resonantly excited by clumps of matter falling into the black hole
at some appropriate rate. This rate should be such that the spatial
separation of the clumps equals a multiple of the typical QNM
wavelength. Here we show by a simple model that, in principle, the
modes of a black hole can indeed be excited in this way.  We also show
that a simple addition of damped sinusoids provides a good fit of the
resulting gravitational waveforms: in other words, we can interpret
the resonant excitation of the modes in terms of interference between
gravitational waves.  For simplicity we consider clumps of mass $\mu$
much smaller than the black hole mass, $\mu\ll M$, so that we can
apply a perturbative analysis.

Consider first one single clump falling from infinity. This process
was first analyzed in a classical paper by Davis, Ruffini, Press and
Price \cite{davis} (hereafter DRPP).  They found that the total
radiated energy is $\simeq 0.0104\mu^2/M$, most of it
($0.0091\mu^2/M$) being emitted as quadrupolar ($\ell=2$) waves, and
that the total energy spectrum is peaked at $\omega \sim 0.32/M$, very
close to the fundamental $\ell=2$ QNM frequency $\omega \simeq
0.3737/M$.

We can now superpose the waveform and its Fourier transform for the
one-particle case to represent two or more particles falling into a
black hole and to study interference phenomena (we are assuming that
the clumps are non-interacting, so that there is no extra contribution
to the total energy-momentum tensor).  A similar
sum-over-point-particles approach has been used many times in the past
\cite{pointptcles}, but the particular process we consider here has
not been studied before.

Let us first suppose that we drop two particles with a temporal
separation of $T$ (which, bearing in mind that the retarded time is
the measured quantity, also means a spatial separation of $T$ between
the two bodies).  If $\Psi_1(t,r)$ represents the waveform for the
first body (normalized by $\mu$, where $\mu$ is the mass of the
infalling body), then the total waveform will be
\be
\Psi(t,r)=\mu \Psi_1(t,r)+\mu \Psi_1(t-T,r)\,.
\ee
Likewise the Fourier transform ${\tilde \Psi}(\omega,r)$ will be
\be
{\tilde \Psi}=\mu {\tilde \Psi_1}+\mu e^{i \omega T}{\tilde \Psi_1}\,.
\ee
For $N$ bodies dropped regularly, such that the temporal difference
between the dropping of the first and of the last is $T$, we have
\be
\Psi(t,r)=\sum_{j=0}^{N-1} \mu \Psi_1\left(t-\frac{i T\,}{N-1}j,r\right)\,,\quad
{\tilde \Psi}(\omega,r)=\sum_{j=0}^{N-1} \mu e^{\frac{i \omega T}{N-1}j} {\tilde \Psi_1}\,.
\ee
The infall of a finite-sized body can also be studied with this
formalism \footnote{
We can think of a finite body as being composed of many particles, the
separation between the particles being infinitesimal. The total
waveform $\Psi$ will be a superposition of single-particle waveforms
$\Psi_1$, of the form
$ \Psi(t,r)=\lim_{N \to \infty} \sum_{j=0}^{N-1} \frac{\mu}{N}
\Psi_1\left(t-\frac{i T}{N-1}j,r\right)\,.$
The Fourier transform of this waveform is
$ {\tilde \Psi}=\lim_{N \to \infty} \sum_{j=0}^{N-1}\f{\mu}{N}
e^{\frac{i \omega T}{N-1}j} {\tilde \Psi_1}\,.  $
By taking this limit we obtain the gravitational radiation generated
by a massive thin body of length $T$ and total mass $\mu$.  Summation
of the series yields $ \lim_{N \to \infty}
\sum_{j=0}^{N-1}\frac{\mu}{N} e^{\frac{i \omega T}{N-1}j}= \mu
\frac{-1+e^{i\omega T }}{i \omega T}\,,$ so that $ {\tilde \Psi}=\mu
\frac{-1+ e^{i \omega T}}{i \omega T} {\tilde \Psi_1}\,.$
}.  Incidentally, one can prove that
$|\mu \sum_{j=0}^{N-1} e^{\frac{i \omega T}{N-1}j}|^2\leq \mu^2 N^2$,
so the total energy $ E_{\rm rad}\sim \omega^2 |{\tilde \Psi}|^2 $
radiated by a stream of plunging particles is always less than or
equal to the energy radiated by a single point particle with total
mass $\mu N$.

\begin{figure}
  \includegraphics[height=.35\textheight,angle=270]{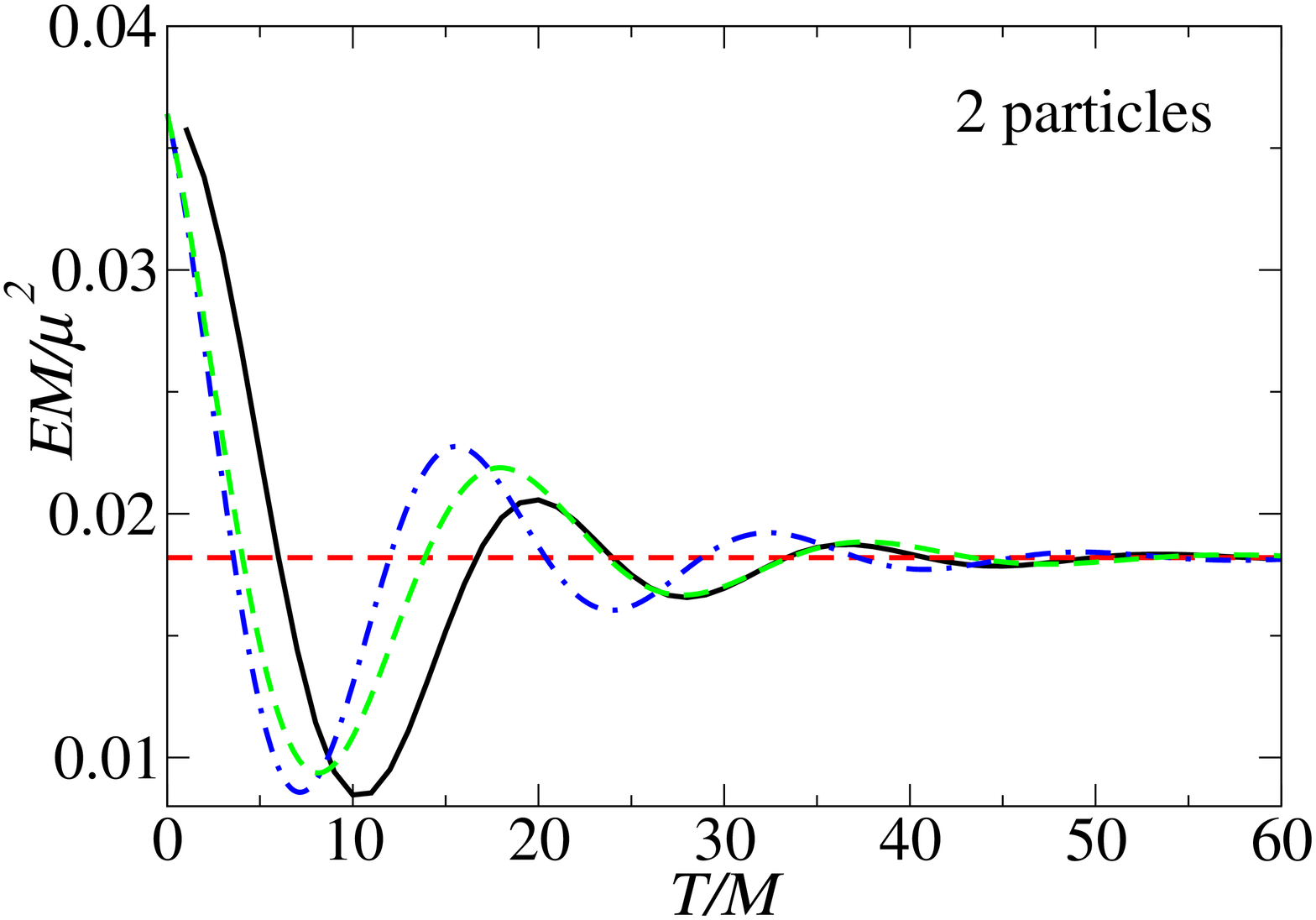}
  \includegraphics[height=.35\textheight,angle=270]{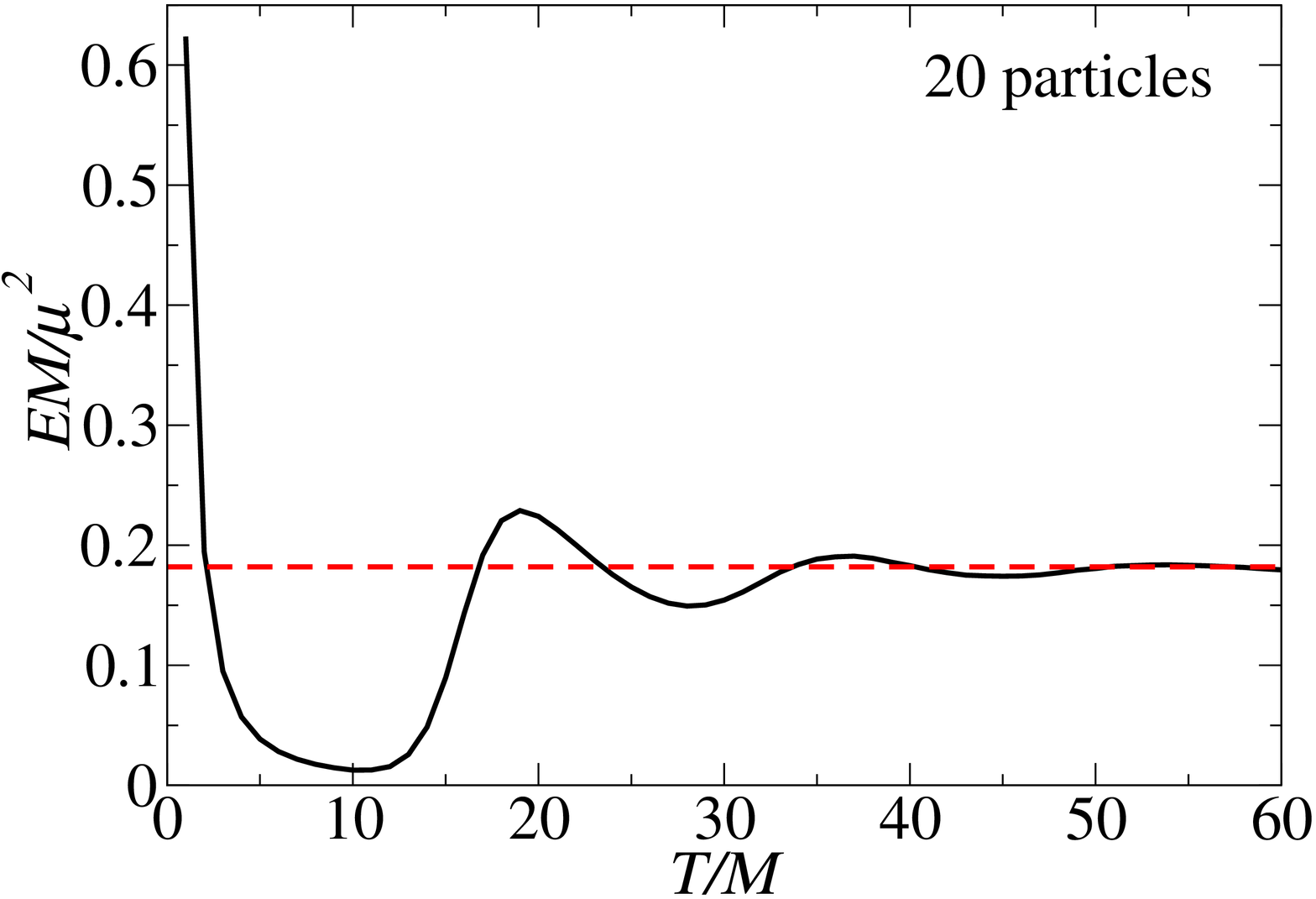}
  \caption{Characteristic interference pattern in the total energy
  radiated by two (left) and twenty (right) particles falling into a
  Schwarzschild black hole, as a function of the particles' initial
  separation. As $T\to 0$ the particles behave as a single particle of
  mass $(N\mu)$, and the total energy scales like $(N\mu)^2$. The red,
  dashed horizontal line corresponds to the sum of the single-particle
  energies ($E\sim N\times 0.0091\mu^2/M$ for $N$ particles). In the
  two-particle case we also overplot the prediction of the two-mode
  ringdown waveform, Eq.~(\ref{twoptclesQN}).  The blue, dot-dashed
  line corresponds to $\omega_r=0.3737/M$ (the fundamental QNM
  frequency with $\ell=2$). To obtain the green, dashed line we choose
  the ``phenomenological'' value $\omega_r=0.32/M$ (corresponding to
  the peak of the one-particle energy spectrum).
\label{fig:powertwotwentyptcles}}
\end{figure}

\begin{figure}
  \includegraphics[height=.35\textheight,angle=270]{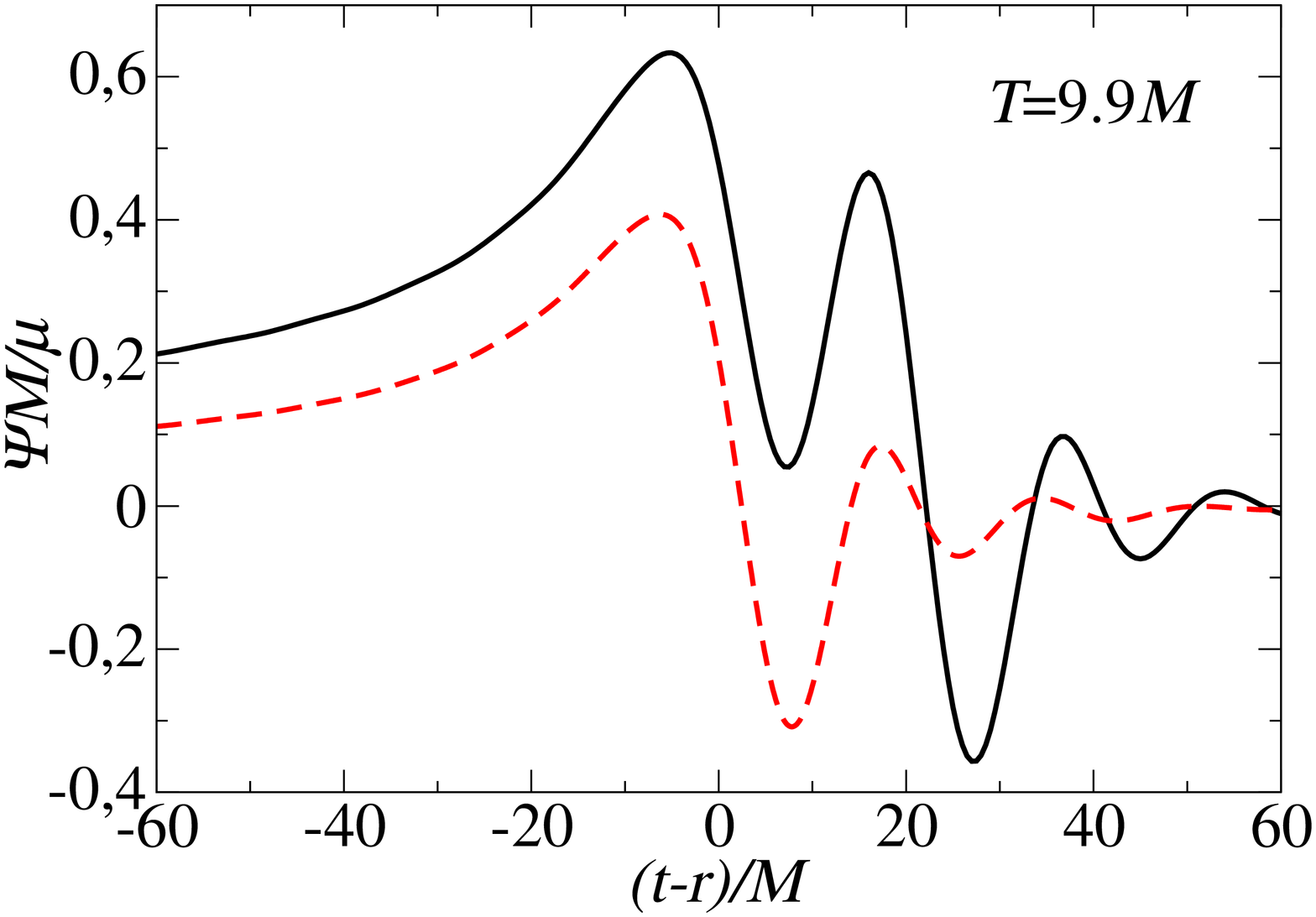}
  \includegraphics[height=.35\textheight,angle=270]{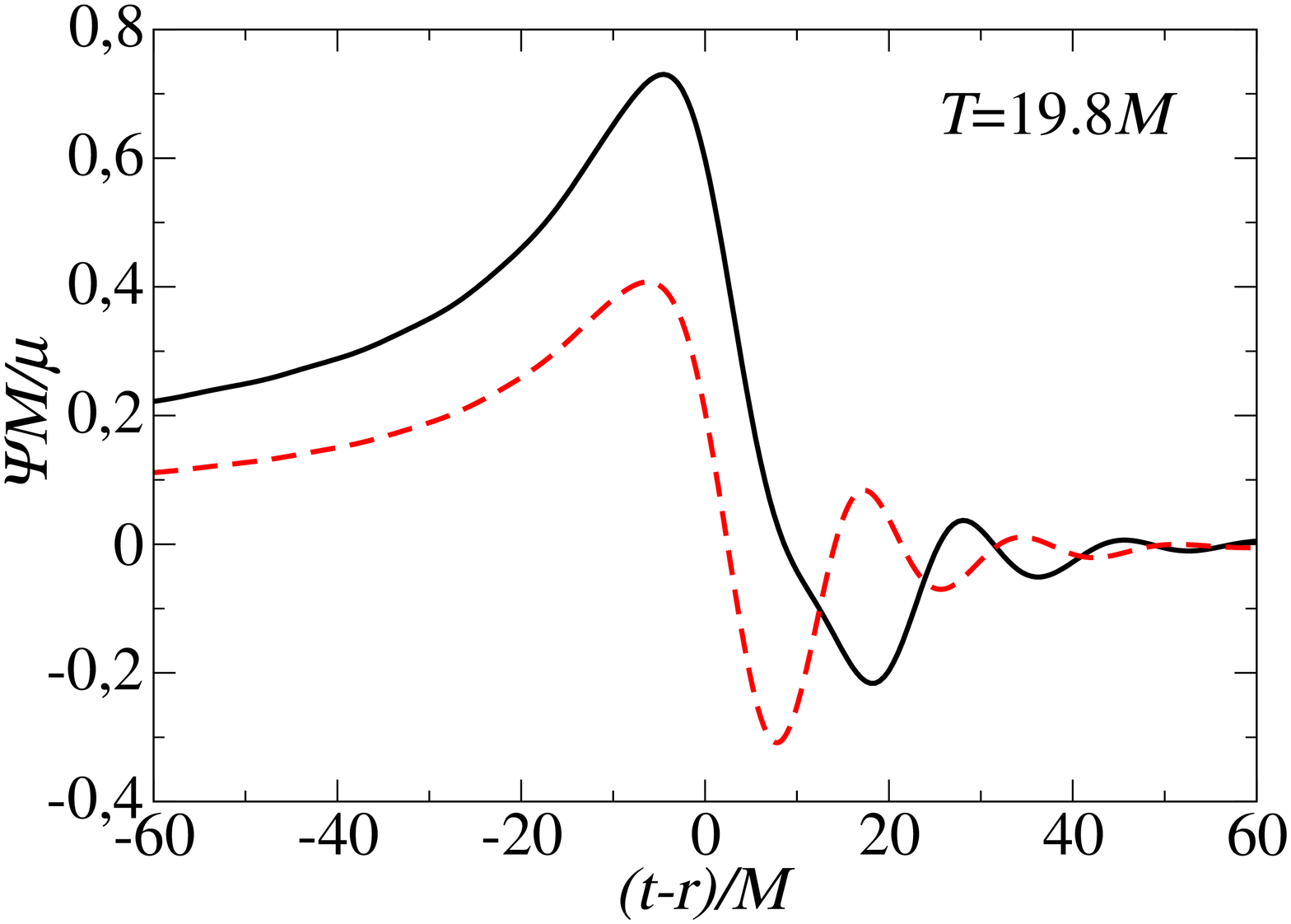}
  \caption{Two-particle waveforms (solid black lines) are compared
  with the single-particle waveforms (dashed red lines). In the left
  panel $T=9.9M$, and the ringdown is suppressed by destructive
  interference. In the right panel $T=19.8M$, and the ringdown is
  enhanced by constructive interference.
\label{fig:waveformstwoptcles}}
\end{figure}

The total energy radiated in $\ell=2$ as a function of separation for
two and twenty particles is shown in
Fig.~\ref{fig:powertwotwentyptcles}. Let us first consider the
two-particle case (left panel).  As we increase the distance between
the two bodies the total energy decreases. At the first minimum
($T\sim 9.9 M$) the total energy radiated is $\sim 0.0085
\mu^2/M$. The next maximum is attained at $T\sim 19.8 M$, the total
radiated energy being $\simeq 0.0206 \mu^2/M$.  Since the one-particle
spectrum is highly peaked at $\omega \sim 0.32/M$, it is tempting to
explain this behavior as an interference phenomenon. The typical
dominant wavelength of the fundamental QNM is $\lambda=2\pi/\omega
\sim 19.6 M$: this means that, for two particles, destructive
interference should occur at $T \sim 9.8 M$ and constructive
interference at $T \sim 19.6 M$, in good agreement with our numerical
results. In the limit of large distances, interference becomes
negligible and the total radiated energy tends to the sum of the
individual particle energies ($\simeq 0.0182 \mu^2/M$ for our two
particles). This interpretation of the result is confirmed by an
explicit calculation of the gravitational waveforms
(Fig.~\ref{fig:waveformstwoptcles}).

The situation for more than two particles is similar. Notice that in
the two-particle case, for $T<6 M$ there is always an enhancement on
the total radiated energy (as compared to the sum of the
single-particle energies).  However, as we increase the number of
particles, this enhancement interval decreases.  So we should expect
it will be hard to resonantly excite the black hole QNMs using a large
number of particles.

\vskip .2cm

%%%%%%%%%%%%%%%%%%%%%%%%%%%%%%%%%%%%%%%%%%%%%%%%%%%%%%%%%%%%%%%%%%%%%%%%%%%%%%%
{\em Interpreting the results in terms of ringdown waveforms - } 
%%%%%%%%%%%%%%%%%%%%%%%%%%%%%%%%%%%%%%%%%%%%%%%%%%%%%%%%%%%%%%%%%%%%%%%%%%%%%%%
The DRPP signal is dominated by quasinormal ringing, so it makes sense
to try and interpret the preceding results in terms of ringdown
waveforms. To start with, we assume our signal has the form
\be
\psi=\Theta(t) Ae^{-\omega_i t}\sin{\omega _r t}\,,
\label{ringdownwaveform}
\ee
where $\Theta(t)$ is the Heaviside function. For the fundamental mode of
a Schwarzschild black hole $\omega _r \simeq 0.3737/M$ and $\omega_i
\simeq 0.0890/M$.  The amplitude $A$ can be obtained imposing that the
total radiated energy in $\ell=2$ is equal to $0.0091 \mu ^2/M$.  The
total energy radiated in the $\ell$-th multipole is
\be
E_{\rm rad}=\frac{1}{64\pi}\frac{(l+2)!}{(l-2)!}
\int_{-\infty}^{-\infty} \left(\frac{d\psi}{dt}\right )^2 dt\,,
\ee
which for the waveform (\ref{ringdownwaveform}) gives
\be
E_{\rm rad}=\frac{A^2(l+2)!\omega_r^2}{256\pi(l-2)!\omega_i}\,.
\label{oneptcleQN}
\ee
Setting $\ell=2$ and equating this to $0.0091\mu^2/M$ gives $A\simeq
0.441 \mu$, in qualitative agreement with the DRPP waveform.  Let us
now consider two particles by superposing two ringdown waveforms:
\be
\psi=\Theta(t) Ae^{-\omega_i t}\sin{\omega _r t}+
\Theta(t-T) Ae^{-\omega_i (t-T)}\sin{\omega _r (t-T)}\,.
\label{ringdownwaveform2ptcles}
\ee
The total energy radiated would be
\be
E_{\rm rad}^{\rm 2
ptcles}=\frac{A^2e^{-T\omega_i}\omega_r(l+2)!}{128\pi(l-2)!\omega_i}
\left (e^{\omega_i T}\omega_r+\omega_r\cos {\omega_r T}-
\omega_i\sin \omega_r T \right )\,.
\label{twoptclesQN}
\ee
A plot of this function is shown in the left panel of
Fig.~\ref{fig:powertwotwentyptcles}. It is in good qualitative
agreement with the full numerical calculation.  If we use a
``phenomenological'' QNM frequency $\omega_r=0.32/M$ (corresponding to
the peak of the one-particle energy spectrum, and yielding an
amplitude $A\simeq 0.515 \mu$) instead of the ``true'' fundamental
$\ell=2$ QNM frequency $\omega_r=0.3737/M$, the agreement between
Eq.~(\ref{twoptclesQN}) and the numerical results becomes even better.

This agreement suggests the possibility to extend our study to Kerr
black holes, by simply assuming that $\omega_r$ and $\omega_i$ are
functions of the black hole's angular momentum.  Using
(\ref{oneptcleQN}) and (\ref{twoptclesQN}) we get
\be
\frac{E_{\rm rad}^{\rm 2 ptcles}}{E_{\rm rad}}=2e^{-T\omega_i}
\left (e^{\omega_i T}+\cos {\omega_r T}-
\frac{\omega_i}{\omega_r}\sin \omega_r T \right )
\sim
2(1+\cos\omega_r T)\,.
\ee
In the last step we considered rapidly rotating (near-extremal) black
holes and modes with $\ell=m$, for which $\omega_i \rightarrow 0$ and
$\omega_r \rightarrow m/(2M)$.  Therefore, for corotating modes of
rapidly rotating black holes the resonance is almost perfect if we
choose $T=2\pi M/m$ (the ratio is then 4, the maximum allowed) while
destructive interference results if $T=\pi M/m$.

In summary: constructive resonance, and consequently QNM excitation,
is possible (in principle) when a stream of particles accretes onto a
black hole. Unfortunately, the separation between each particle
required for an enhancement in the total energy becomes very small as
the number of particles increases.  Unless some delicate fine tuning
is at work \cite{AG}, the separation between any two infalling
particles or clumps should be driven by a random process, which means
that on average there should be no resonance at all. This is confirmed
by simulations of generic configurations of infalling matter, where
quasinormal ringing is usually absent from the waveforms \cite{nagar}.

%%%%%%%%%%%%%%%%%%%%%%%%%%%%%%%%%%%%%%%%%%%%%%%%%%%%%%%%%%%%%%%%%%%%%%%%%%%%%%%
\section{Energy distribution in different scenarios}
%%%%%%%%%%%%%%%%%%%%%%%%%%%%%%%%%%%%%%%%%%%%%%%%%%%%%%%%%%%%%%%%%%%%%%%%%%%%%%%

In this Section we present a brief review of results on QNM excitation
in different physical scenarios, including gravitational collapse,
evolutions of single distorted black holes and head-on black hole
collisions. Far from being complete or comprehensive, our review aims
to emphasize areas that require further investigation from the point
of view of gravitational wave source modeling.

%%%%%%%%%%%%%%%%%%%%%%%%%%%%%%%%%%%%%%%%%%%%%%%%%%%%%%%%%%%%%%%%%%%%%%%%%%%%%%%
{\em Gravitational collapse - }
%%%%%%%%%%%%%%%%%%%%%%%%%%%%%%%%%%%%%%%%%%%%%%%%%%%%%%%%%%%%%%%%%%%%%%%%%%%%%%%
There are just a few calculations of gravitational wave emission from
rotating gravitational collapse to a black hole. Perturbative
calculations were first carried out by Cunningham, Price and Moncrief
\cite{cpm1,cpm2,cpm3}, and later improved upon by Seidel and
collaborators \cite{seidel}. These studies suggest that gravitational
waves are mainly generated in the region where the Zerilli (or
Regge-Wheeler) potential is large, and that the signal is always
dominated by quasinormal ringing of the finally formed black
hole. Simplified simulations based on a free-fall (Oppenheimer-Snyder)
collapse model yield a rather pessimistic energy output, with a
typical core-collapse radiating up to $E\simeq 10^{-7} M$ in
gravitational wave energy, and most of the radiation being emitted in
modes with $\ell=2$. Radiation in $\ell=3$ is typically two to three
orders of magnitude smaller than the $\ell=2$ radiation (see
eg. Fig. 9 in \cite{cpm1}). Cunningham {\it et al.} also provide
analytical expressions for the energy radiated in different multipoles
as a function of the initial quadrupole deformation of the star and of
the initial collapse radius \cite{cpm1,cpm2}.

For a long time the only {\it nonperturbative}, axisymmetric
calculation of gravitational wave emission from collapse has been
provided by the seminal work of Stark and Piran
\cite{starkpiran}. Surprisingly, the waveform resembles that emitted
by a point particle falling into a black hole, but with a reduced
amplitude (we will see below that a similar result emerges from
head-on black hole collisions in full general relativity). As expected
for axisymmetric configurations, the amplitude of the cross
polarization mode is always $<0.2$ that of the even mode, and the
emission in the plus mode is dominantly quadrupolar, with a $\sin^2
\theta$ angular dependence (the cross polarization has a dependence of
the form $\cos\theta \sin^2\theta$). The total energy emitted is quite
low; it increases with rotation rate, ranging from $\sim 10^{-8} M$
for $j=0$ to $\sim 7\times 10^{-4} M$ as $j\to 1$. Rotational effects
halt the collapse for some critical value of $j=j_{\rm crit}$ which is
very close to unity, and depends on the amount of artificial pressure
reduction used to trigger the collapse. Stark and Piran find that the
energy emitted scales as $j^4$ for low $j$: $E/M\simeq {\rm
min}\{1.4\times 10^{-3}j^4,\epsilon_{\rm max}\}$ for $0<j<j_{\rm
crit}$, where $\epsilon_{\rm max}\sim 10^{-4}$ again depends on the
amount of artificial pressure reduction used to trigger the
collapse. The $h_+$ waveform is very well fitted by a combination of
the two lowest QNMs with $\ell=2$ and it is only weakly sensitive to
$j$, reflecting the weak dependence of the slowly-damped Kerr QNM
frequencies on $j$.

This two-dimensional calculation has recently been improved upon using
a three-dimensional code \cite{baiotti}, but still keeping a high
degree of axisymmetry (so only modes with $m=0$ are
excited). Ref.~\cite{baiotti} picks as the initial configuration the
most rapidly rotating, dynamically unstable model described by a
polytropic EOS with $\Gamma=2$ and $K=100$, having a dimensionless
rotation rate $\simeq 0.54$, and triggers collapse reducing the
pressure by less than about $2\%$.  Gravitational waves are extracted
by a gauge-invariant approach in which the spacetime is matched with
nonspherical perturbations of a Schwarzschild black hole. The ``plus''
polarization is essentially a superposition of modes with $\ell=2$ and
$\ell=4$, and the ``cross'' polarization is a superposition of modes
with $\ell=3$ and $\ell=5$. Fig.~1 in \cite{baiotti} shows that the
amplitude of the $\ell=2$ mode is about an order of magnitude larger
than the amplitude of the $\ell=4$ mode, and Fig. 3 shows that the
cross polarization is roughly suppressed by one order of magnitude
with respect to the plus polarization, with maximum amplitudes in a
ratio $|(r/M) h_\times|_{\rm max}/|(r/M) h_+|_{\rm max}\simeq 0.06$
(here $r$ denotes the distance between source and
detector). Odd-parity perturbations should be zero in spacetimes with
axial and equatorial symmetries: they are nonzero because of the
rotationally induced coupling with even-parity perturbations
\cite{cpm1,cpm2,cpm3,seidel}. The maximum amplitude from these
three-dimensional simulations is $|(r/M) h_+|_{\rm max}\simeq
0.00225$, about one order of magnitude smaller than the amplitude
found by Stark and Piran. Correspondingly, the total energy lost to
gravitational waves is $E\simeq 1.45\times 10^{-6} (M/M_\odot)$. This
is about two orders of magnitude smaller than the estimate by Stark
and Piran for the same value of the angular momentum, but larger than
the energy losses found in recent calculations of rotating stellar
core collapse to protoneutron stars \cite{muller}.

\vskip .2cm

%%%%%%%%%%%%%%%%%%%%%%%%%%%%%%%%%%%%%%%%%%%%%%%%%%%%%%%%%%%%%%%%%%%%%%%%%
{\em Distorted black holes - }
%%%%%%%%%%%%%%%%%%%%%%%%%%%%%%%%%%%%%%%%%%%%%%%%%%%%%%%%%%%%%%%%%%%%%%%%%
Distorted black hole simulations shed some light on the (crucial)
issue of the dependence of QNM excitation on the initial data: that
is, on the ``shape'' and ``size'' of the black hole distortion.

Ref.~\cite{krivan} carries out a pioneering study of linearized
gravitational perturbations of the Kerr spacetime. Given initial data
with a certain angular dependence - that is, assigned values of
$(l,~m)$ - their study finds that modes with positive and negative $m$
are always excited together. The simultaneous excitation of corotating
and counterrotating modes depends on a certain reflection-symmetry
property of the Kerr QNM spectrum \cite{BCW}, and is confirmed by more
recent time-evolutions of scalar perturbations of the Kerr spacetime
\cite{dorband}. In addition, these new simulations show that the
relative excitation of corotating and counterrotating modes depends on
the radial profile of the initial data, confirming expectations based
on analytic investigations of the same problem
\cite{GA}. Unfortunately, detailed studies of the excitation of
gravitational perturbations of Kerr black holes in the perturbative
regime are still lacking.

Ref.~\cite{abrahams92} provides one of the first attempts to extract
gravitational waveforms from {\it non-linear} simulations of distorted
black holes. They use a two-dimensional code to evolve initial data
corresponding to a single, nonrotating black hole superimposed with
time-symmetric gravitational waves (Brill waves), extract
gravitational waves using gauge-invariant perturbation theory and show
that, for low-amplitude Brill waves, non-perturbative and perturbative
waveforms with $\ell=2$ and $\ell=4$ are in good agreement.

A more complete analysis is due to Anninos {\it et al.}
\cite{bernstein2}. They use three different initial data sets,
corresponding to very different physical situations: non-rotating
black holes distorted by time-symmetric Brill waves, distorted
rotating black holes, and the time-symmetric ``two black hole'' Misner
data. In all three cases the system quickly settles down to a nearly
spherical (or oblate, for rotating holes) configuration. They study
the dynamics of the black hole horizon monitoring the ratio of polar
and equatorial circumference $C_r=C_p/C_e$, and find that the horizon
oscillates at the QNM frequencies of the final black
hole. Oscillations with $\ell=2$ and $\ell=4$ are well fitted by a
superposition of the fundamental QNM and the first overtone. In the
``black hole plus (small) Brill wave'' case, even when the initial
data contain a significant component with $\ell=4$ (that is, when the
index $n$ in the analytic expression of the Brill wave is equal to 4)
the $\ell=4$ component of the horizon distortion is a factor $\sim
10^{-4}$ smaller than the $\ell=2$ component (Figs. 2 and 3 in
\cite{bernstein2}). For Misner initial data, corresponding to
colliding black holes, $\ell=2$ waves are also strongly dominant. An
important conclusion of Ref.~\cite{bernstein2} is that the dynamics of
the apparent horizon geometry can be used not only to find the
fundamental QNM frequency of the hole, but also its mass and angular
momentum.

Brandt and Seidel \cite{BS} present an extensive study of distorted,
rotating ``Kerr-like'' black holes with a wide range of rotation
parameters (as high as $j\simeq 0.87$). Quasinormal ringing is
observed both in the even and in the odd components of the emitted
radiation. The angular momentum parameter $j$ is well approximated by
the simple fit $j=\sqrt{1-(-1.55+2.55 C_r)^2}$ (the fit being
generally accurate within $\sim 2.5\%$): this means that a measurement
of the horizon geometry provides a value of the rotation parameter.
For black holes with $j>\sqrt{3}/2$, a Kerr black hole cannot be
embedded in flat space: an alternative method to obtain $j$ consists
in measuring the angle at which the embedding ceases to exist.  

A crucial result of Ref.~\cite{BS} is that the black hole dynamics
depend critically on the properties of the initial data sets. For
example, initial data corresponding to nonlinear, odd perturbations
produce oscillations of the Gaussian curvature at frequencies given by
linear combinations of QNMs with $\ell=3$ and $\ell=5$, because of
beating phenomena. In Sec.~IVB, they present a method to ``dig out''
frequencies with stronger damping from the slowly-damped modes in a
Fourier transform, which can find useful applications in
gravitational-wave data analysis. Sec.~V of~\cite{BS} presents
waveforms corresponding to different initial distortions of the
hole. Odd-parity distortions of Schwarzschild black holes (run ``o1''
in their terminology) produce a waveform in which $99.9\%$ of the
energy is radiated in $\ell=3$, $0.05\%$ of the radiation goes into
$\ell=2$, $0.02\%$ into $\ell=4$ and $0.0001\%$ into $\ell=5$; but for
different ``shapes'' of the distortion (run ``o2'') , the energy
radiated in $\ell=5$ can be as large as $21\%$. Similarly, runs
corresponding to distorted Kerr black holes (``r0'' to ``r5'') contain
different mixtures of the different multipoles depending on the shape
of the initial distortion and on the rotation rate of the black
hole. As a trend, large rotation seems to yield a large $\ell=3$
component: the maximum corresponds to run ``r4'', for which the
$\ell=2$ component carries $92.1\%$ of the energy and $\ell=3$ carries
$\sim 7\%$ of the energy.

Three-dimensional, non-axisymmetric evolutions are urgently needed. A
handful of non-axisymmetric simulations can be found in
Ref.~\cite{allen}. They find that the total energy radiated in a given
multipole, $E_{\ell m}$, ranges from $E_{20}\sim 3\times 10^{-4} M$ to
$E_{42}\sim 3\times 10^{-7} M$. Only modes with $\ell=2$, $m=0,~2$ and
$\ell=4$, $m=0,~2$ should occur at linear order in the amplitude, all
other modes being quadratic in the amplitude (for one of these
``nonlinear'' modes they find a completely negligible energy,
$E_{62}\sim 10^{-10}M$).

In fact, one of the most interesting features introduced by
nonlinearities is mode-mode coupling. Ref.~\cite{zlochower} studies
this coupling in detail for nonrotating (Schwarzschild) black
holes. Large-amplitude incident waves effectively add mass to the
black hole, producing a small but visible ``redshift'' in the QNM
frequencies. The waveform is also slightly shifted by
nonlinearities. This nonlinear dephasing might be an important issue
when applying matched-filtering techniques. On the other hand, when
the amplitude $A$ of incident waves exceeds a certain threshold,
nonlinearities {\it amplify} the outgoing wave, which of course is
good news for detection. Beyond the ``nonlinearity threshold'' the
outgoing amplitude of the $\ell$-th multipole scales as
$A^{\ell/2}$. The reason is that, for ordinary spherical harmonics,
$\left(Y_{\ell m}\right)^k\sim Y_{(k\ell)m}+$terms with smaller
$\ell$. Given a linear incident wave with $\ell=2$, an $\ell=2k$ mode
arises from order $k$ (and higher) nonlinear terms; this mode's
amplitude scales as $A^k+{\cal O}(A^{k+1})$. So, the quadratic
coupling of an $(\ell=2,~m=0)$ mode with itself produces an
$(\ell=4,~m=0)$ mode of order $A^2+{\cal O}(A^3)$ (and higher), in
addition to an $(\ell=2,~m=0)$ mode of order $A+{\cal O}(A^2)$ (notice
that $\ell=0$ is ruled out for gravitational perturbations, being
non-radiative). In addition, nonlinearities produce beating
phenomena. In general, quadratic terms of an harmonic with frequency
$\omega$ contain the frequencies $(0,~2\omega)$, cubic terms contain
$(\omega,~3\omega)$, and so on. The selection rules for mode coupling
in Kerr backgrounds should be more interesting: even in the linear
regime, the Lense-Thirring effect couples axial and polar
perturbations of a slowly rotating star according to Laporte's rule
\cite{laporte}. Mode couplings in the Kerr background can be very
important for gravitational wave data analysis, and deserve further
study.

In conclusion, simulations suggest that (at least in situations with a
certain degree of symmetry) high overtones are heavily
suppressed. Investigating an extensive catalogue of initial data,
Ref.~\cite{BS} shows that we just {\it don't know} how energy is going
to be partitioned between different modes in a realistic merger event,
unless we can predict with accuracy the shape of the initial
data. Nonetheless, for data-analysis purposes it should be reasonable
to assume that modes with $\ell=2$, $m=0,~\pm 2$ have the largest
amplitude (the components with $|m|=2$ probably being dominant). The
contribution from $\ell=4$, $m=0,~\pm 2$ could be smaller by about two
orders of magnitude in energy. Modes with $\ell=3$ may also be
relevant, especially at large rotation rates. Nonlinearities could
produce a cascade of energy into higher-order modes, but we need more
simulations with realistic initial data to assess the relevance of
this effect for QNM detection.

\vskip .2cm

%%%%%%%%%%%%%%%%%%%%%%%%%%%%%%%%%%%%%%%%%%%%%%%%%%%%%%%%%%%%%%%%%%%%%%%%%%%%%%%
{\em Head-on black hole collisions - }
%%%%%%%%%%%%%%%%%%%%%%%%%%%%%%%%%%%%%%%%%%%%%%%%%%%%%%%%%%%%%%%%%%%%%%%%%%%%%%%

In four dimensions, the total energy radiated by a freely-falling
particle released from rest at infinity in the Schwarzschild spacetime
\cite{davis,ferrari} is in surprising agreement with numerical
simulations of head-on black hole collisions.  An extensive comparison
of numerical results with perturbation theory can be found in
Refs.~\cite{smarr,anninos1,anninos2,closelimit,anninosl4}.

Fig.~14 and Sec.~IV of Ref.~\cite{anninos2} show that the
perturbation-theory prediction for the radiated energy, $E=0.0104
\mu^2/M$~\cite{davis}, is in excellent agreement with numerical
results when we replace the particle's mass by the system's reduced
mass. In fact, perturbation theory slightly {\it overestimates} the
energy output predicted by numerical simulations, which (according to
state-of-the-art simulations) is $E\simeq 0.0013 M$
\cite{sperhake}. In the so-called ``particle-membrane'' picture, this
disagreement is compensated for multiplying the perturbative
prediction by three fudge factors~\cite{anninos2}: i) a factor
$F_{r_0}$, accounting for the fact that in numerical simulations the
infall starts at finite distance $r_0$; ii) a factor $F_h$ coming in
because the black hole (unlike the falling particle) has finite size,
and tidal deformations heat up the horizon; iii) a factor $F_{\rm
abs}$ to account for reabsorption of the emitted gravitational waves
by each black hole. If the collision starts at infinite separation,
$F_{r_0}=1$; $F_h$ has a weak dependence on $r_0$, and tends to
$F_h\simeq 0.86$ as $r_0\to \infty$; $F_{\rm abs}\simeq 0.99$ is even
less relevant.

If the total energy output from head-on black hole collisions is well
understood, unfortunately the relative excitation of higher multipoles
with respect to $\ell=2$ is not. Ref.~\cite{anninosl4} compares
numerical calculations with the close-limit approximation (valid when
the black holes are close enough to be considered as perturbations of
a single Schwarzschild black hole) and with the particle-membrane
approach.  Fig.~12 of Ref.~\cite{anninosl4} compares the radiated
energy according to the close-limit approximation with results from
the numerical simulations. Extraction of the $\ell=4$ component from
the simulations is very sensitive to numerical noise, especially when
the black holes are very close and the $\ell=4$ component is more
heavily suppressed.

Numerical simulations have been extended to include unequal-mass black
holes \cite{brandt} and boosted black holes
\cite{boosted,boosted2}. More recently, mesh refinement techniques
were used to obtain more reliable estimates of the energy radiated in
$\ell=2$ \cite{sperhake}.  Even for highly boosted black holes, the
agreement between perturbation theory (in the close-limit
approximation) and numerical simulations is very good.  Numerical
studies of boosted black holes show that, fixing the initial
separation between the holes, the energy radiated saturates at $E\sim
0.01 M$ for very large values of the initial momentum of the holes
(see eg. Fig.~2 of \cite{boosted2}). Ref.~\cite{boosted} combines
Newtonian dynamics and numerical simulations of boosted black holes to
suggest that the maximum energy emission could actually be much lower
than this, $E<0.0016 M$. In general, these studies suggest that the
role of the weak-field phase of the evolution is only to determine the
momentum of the black holes when they start to interact nonlinearly,
emitting most of the radiation in the ringdown phase. This expectation
is confirmed by the comparison of perturbation theory and
Post-Newtonian calculations \cite{simone}: the bremsstrahlung
radiation emitted by a particle at separations larger than $4M$ (in
Schwarzschild coordinates) contributes only $\sim 3 \%$ of the total
energy, most of the radiation coming from the final (ringdown) phase.

\section{Conclusions}

Most numerical simulations so far consider nearly-axisymmetric
situations, so they don't have much to say about the distribution of
energy between modes with different values of $(l,m)$ in a realistic
black hole merger. This is a crucial issue for data analysis from
ringdown waveforms \cite{BCW}. Recently there has been some remarkable
progress in non-axisymmetric simulations of black hole mergers (see
eg. \cite{herrmann} and references therein), but a discussion of these
developments is beyond the scope of this paper.

Some estimates of the energy radiated in the plunge phase by
solar-mass binaries of rotating black holes are provided by the
effective-one-body approach \cite{EOB}. The error introduced by
extrapolating to supermassive black holes should not be too large,
since the energy balance of black hole binaries involves a single
fundamental scale (their total mass $M$). It is useful to compare the
effective-one-body estimate with earlier estimates of the energy
radiated in the plunge {\it plus ringdown} phases, as provided by the
Lazarus approach \cite{Lazarus}. For aligned (anti-aligned) spins and
specific angular momenta $j_1=j_2=0.17$ ($j_1=j_2=0.25$) the
effective-one-body approach predicts an energy release $\sim
(0.6-0.9)\% M$ [$\sim (1-3)\% M$]. The Lazarus approach finds $\sim
(1.7-1.9)\% M$ [$\sim (1.9-2.1)\% M$], respectively. In principle, the
difference of these two estimates should bracket the most reasonable
range of ringdown efficiencies. However, the overall error on both
estimates is still too large to draw any definite
conclusion. Numerical relativity should be able to provide a reliable
answer to this problem within a very short time.

\begin{theacknowledgments}
We thank Leonardo Gualtieri, Luis Lehner and Alessandro Nagar for
useful discussions. This work was supported in part by the National
Science Foundation under grant PHY 03-53180.
\end{theacknowledgments}



\bibliographystyle{aipproc}   % if natbib is available

%%%%%%%%%%%%%%%%%%%%%%%%%%%%%%%%%%%%%%%%%%%
%% The following lines show an example how to produce a bibliography
%% without the help of the BibTeX program. This could be used instead
%% of the above.
%%%%%%%%%%%%%%%%%%%%%%%%%%%%%%%%%%%%%%%%%%%

%\endinput

\end{document}